\documentclass{article}%
\usepackage{amssymb}
\usepackage{epsfig}
\usepackage{psfig}
\usepackage{latexsym}%
\usepackage{amsmath}%
\usepackage{amsfonts}%
\usepackage{graphicx}
\begin{document}
\begin{flushright}
October 2004 \\
\end{flushright}

\medskip

\begin{center}
{\LARGE $B$ and $B_{s}$ decay constants from QCD Duality at three
loops\footnote{{\LARGE {\footnotesize Supported by MCYT-FEDER under contract
FPA2002-00612, and partnership Mainz-Valencia Universities.}}}}\vspace{2cm}

{\large J. Bordes}$^{a}${\large , J. Pe\~{n}arrocha}$^{a}${\large \ and K.
Schilcher}$^{b}\bigskip$

$^{a}${\normalsize Departamento de F\'{\i}sica Te\'{o}rica-IFIC, Universitat
de Valencia}

{\normalsize E-46100 Burjassot-Valencia, Spain}

$^{b}${\normalsize Institut f\"{u}r Physik,
Johannes-Gutenberg-Universit\"{a}t}

{\normalsize D-55099 Mainz, Germany\vspace{2cm}}

\textbf{Abstract\medskip}
\end{center}

\begin{quote}
Using special linear combinations of finite energy sum rules which minimize
the contribution of the unknown continuum spectral function, we compute the
decay constants of the pseudoscalar mesons $B$ and $B_{s}$. In the
computation, we employ the recent three loop calculation of the pseudoscalar
two-point function expanded in powers of the running bottom quark mass. The
sum rules show remarkable stability over a wide range of the upper limit of
the finite energy integration. We obtain the following results for the
pseudoscalar decay constants: $f_{B}=178 \pm 14\ MeV$ and $f_{B_{s}}%
=200 \pm 14\ MeV$. The results are somewhat lower than recent predictions based
on Borel transform, lattice computations or HQET. Our sum rule approach of
exploiting QCD quark hadron duality differs significantly from the usual ones,
and we believe that the errors due to theoretical uncertainties are smaller.

\bigskip
\end{quote}

PACS: 12.38.Bx, 12.38.Lg. {\newpage}

\vspace*{2cm}

\section{Introduction}

The decay constant of a pseudoscalar meson $B_{q}$ consisting of a heavy
$b$-quark and a light quark $q$, with $q=u,d,s$, is defined through the matrix
element of the pseudoscalar current:
\[
<\Omega|\,(m_{b}+m_{q})\,\overline{q}\,i\,\gamma_{5}\,b)(0)\,|B_{q}%
>\,=\,\,f_{B_{q}}\,M_{B_{q}}^{2}.
\]

These decay constants are of great phenomenological interest since they enter
in the input to non-leptonic $B$-decays, in the hadronic matrix elements of
$B-\bar{B}$ mixing and in the extraction of $|V_{cb}|$ from the leptonic decay
widths of $B$-Mesons. Knowledge of the decay constants allows to estimate the
so-called hadronic $B$ parameter which is directly related to the deviation of
the vacuum saturation hypothesis. The decay constants are therefore of central
interest to the ongoing experiments carried out in B-factories. Unfortunately
these matrix elements could not be measured directly so far, so that we have
to rely on theoretical calculations. As the calculations must be
non-perturbative there are essentially two approaches, QCD sum rules and
lattice simulations.

The method of the sum rules has been successfully applied since the pioneering
work of Shifman, Vainshtein and Zacharov \cite{SHIFMAN} to calculate various
low energy parameters in QCD. Particular sum rules are based on Borel, Hilbert
transforms, positive moments or inverse moments. Sum rule calculations of the
decay constants have been performed since the eighties using Borel transform
techniques with results within the range $f_{B}=160-210\,\,\mathrm{MeV}$ and
$f_{B_{s}}/f_{B}=1.09-1.22$,
\cite{REINDERS,NARISON,DOMINGUEZPAVER,NEUBERT,RUECKL,BAGAN}. A more recent sum
rule calculations using the new $O(\alpha_{s}^{2})$ correlation function with
one heavy and one light current \cite{JAMINLANGE} yields higher central values
$f_{B}=210\pm19\,\mathrm{MeV}$ and $f_{B_{s}}/f_{B}=1.16\pm0.04$. Lattice QCD
determinations also give results in a wide range $f_{B}=161-218\,\mathrm{MeV}$
and $f_{B_{s}}/f_{B}=1.11-1.16$ \cite{APE,UKQCD} (for a review and a
collection of the results, see \cite{COLANGELO}). A recent HQET calculation
\cite{PS} yields $f_{B}=206\pm20\,\mathrm{MeV}$. The large variation of the
quoted values indicates that there is still room for improvement. 

In general, the sum rule technique assumes duality. In our analysis duality
means that weighted integrals over experimental measured amplitudes should
agree with the same integrals evaluated in QCD perturbation theory. 

We use a method based on finite energy sum rules (FESR) which equates positive
moments of data and QCD theory to evaluate of the decay constants of the
pseudoscalar bottom mesons ($f_{B}$ and $f_{B_{s}}$). The method we propose
here employs a particular combination of positive moments of FESR in order to
optimize the effect of the available experimental information. On the
theoretical side it uses the asymptotic (large momentum expansion) of QCD,
i.e. an expansion in $m_{b}^{2}/s$, where $m_{b}$ is the mass of the bottom
quark and $s$ the square of the CM energy. Such an expansion makes sense as
long as $s$ is far enough from the continuum threshold. We will consider the
perturbative expansion up to second order in the strong coupling constant and
up to fourteen powers in the expansion of the b-mass over the energy, using
the results of the reference \cite{CHETYRKINSTEINHAUSER}. On the
phenomenological side of the sum rule we will consider the lowest lying
pseudoscalar $B_{q}$-meson. In our method we circumvent the problem of the
unknown continuum data by a judicious use of quark-hadron duality. We use a
linear combination of finite energy sum rules of positive moments, designed in
such a way that the contribution of the data in the region above the
resonances turns out to be practically negligible \cite{ACD}.

The present paper differs from our earlier pilot work \cite{BORDES1} in that
$O(\alpha_{s})^{2}$ corrections to the perturbative piece and the
$O(\alpha_{s})$ corrections to the leading non-perturbative term are included.
In this way the stability of the prediction of $f_{B}$ is improved and the
errors are reduced. The stability analysis is also placed on more solid footing.

The plan of this note is the following: in the next section we briefly review
the theoretical method proposed, in the third section we discuss the
theoretical and experimental inputs used in the calculation and in the fourth
one we write up the conclusions. Finally in the appendix we discuss briefly
the main issues of the polynomials used in the finite energy sum rule.

\section{The method}

The two-point function relevant to our problem is:
\[
\Pi(s=q^{2})\,=\,i\int\,dx\,e^{iqx}<\Omega|\,T(j_{5}(x)\,j_{5}(0))\,|\Omega>,
\]
where $<\Omega\,|$ is the physical vacuum and the current $j_{5}(x)$ is the
divergence of the axial-vector current:
\[
j_{5}(x)\,=\,(M_{Q}+m_{q}):\overline{q}(x)\,i\,\gamma_{5}\,Q(x):
\]
$M_{Q}$ is the mass of the heavy bottom quark $Q(x)$, and $m_{q}$ stands for
the mass of the light quarks $q(x)$, up, down or strange. The starting point
of our sum rules is Cauchy's theorem applied to the two-point correlation
function $\Pi(s)$, weighted with a polynomial $P(s)$
\begin{equation}
\frac{1}{2\pi i}\oint_{\Gamma}\,P(s)\,\,\Pi(s)\,\,ds\,\,=\,\,0.\label{CAUCHY}%
\end{equation}

The integration path extends over a circle of radius $\left\vert s\right\vert
=s_{0}$, and along both sides of the physical cut $s\,\in\,\left[
s_{\mathrm{phys.}},s_{0}\right]  $ where $s_{\text{phys.}}$ is the physical
threshold. Neither the polynomial $P(s)$ nor the power of the integration
variable change the analytical properties of $\Pi(s)$, so that we obtain the
following sum rule:
\begin{equation}
\frac{1}{\pi}\int_{s_{\mathrm{phys.}}}^{s_{0}}P(s)\,\mathrm{Im}\,\Pi
(s)=-\frac{1}{2\pi i}\oint_{\left\vert s\right\vert =s_{0}}P(s)\,\,\Pi
(s)\,\,ds,\label{SR}%
\end{equation}

Duality now means that on the left hand side we can use experimental
information starting from the physical threshold $s_{\mathrm{phys.}}$ to $s_{0}%
$, whereas on the right hand side, i. e. on a circle of radius $s_{0}$, we can
use the theoretical input given by QCD and the operator product expansion. The
integration radius $s_{0}$ has to be chosen large enough so that the
asymptotic expansion $\Pi^{\mathrm{QCD}}(s)$ of QCD, which includes
perturbative and non-perturbative terms, constitutes a good approximation to
the two-point correlator on the circle.

On the left hand side of equation (\ref{SR}), we can parametrize the
absorptive part of the two-point function by means of a single resonance
$B_{q}$ and the hadronic continuum of the $b\overline{q}$ channel from
$s^{\mathrm{phys.}}_{\text{cont.}}$ with $s^{\mathrm{phys.}}_{\mathrm{cont.}}\geqslant s_{\mathrm{phys.}}$.
Therefore, we write the representation of the hadronic experimental data in
the form:
\begin{equation}
\frac{1}{\pi}\,\,\mathrm{Im}\,\Pi(s)\,\,\,=\,\,\,\frac{1}{\pi}\,\,\mathrm{Im}%
\,\Pi^{\mathrm{res.}}(s)\,\,+\,\,\frac{1}{\pi}\,\,\mathrm{Im}\,\Pi
^{\mathrm{cont.}}_{\mathrm{phys.}}\,\,\theta(s-s^{\mathrm{phys.}}_{\mathrm{cont.}}), \label{2EXP}%
\end{equation}
where $s^{\mathrm{phys.}}_{\mathrm{cont.}}$ stands for the physical threshold of the
continuum physical region, and therefore it is an input in our calculation. It is worth 
mentioning that it must not be confused with the duality parameter ($s_0$) that
is determined by stability requirements in most versions of finite energy QCD sum rules.

On the right hand side of equation (\ref{SR}) the theoretical input
necessary to write down the two-point function relevant to the integration
over the circle of radius $s_{0}$ has both perturbative and non-perturbative contributions,%

\begin{equation}
\Pi^{\mathrm{QCD}}(s)\,=\,\Pi^{\mathrm{pert.}}(s)\,+\,\Pi^{\mathrm{nonpert.}%
}(s),\label{2QCD}%
\end{equation}
For the perturbative piece, we take the two-point correlation function
$\Pi^{\mathrm{pert.}}(s)$ which has been calculated in
\cite{CHETYRKINSTEINHAUSER} for one massless and one heavy quark as an
expansion up to second order (three loops) in the strong coupling constant
$\alpha_{s} $ and as a power series in the pole mass of the heavy quark
($M_{b}^{2}/s$) up to the seventh order. We have checked that for the employed
range of values of the radius $s_{0}$ of the integration contour, the power
series converges well and does not introduce any appreciable error in the
calculation. In the case of one and two loops the known complete analytic
expressions of the two-point function could be used, but we also use the mass
expansion in this case because the results of the integration can be given the
analytically. Numerically it makes no difference whether one uses the complete
analytic expressions or the expansions.

The authors of \cite{CHETYRKINSTEINHAUSER} obtain the following compact
expansion of the two-point function in terms of the pole mass $M_{b}$,
\begin{equation}
\Pi^{\mathrm{pert.}}(s)=\Pi^{(0)}(s)+\left(  \frac{\alpha_{s}(M_{b})}{\pi
}\right)  \Pi^{(1)}(s)+\left(  \frac{\alpha_{s}(M_{b})}{\pi}\right)  ^{2}%
\Pi^{(2)}(s),\label{PERT1}%
\end{equation}
where the different terms of the expansion in $\alpha_{s}$ have the form:
\begin{equation}
\Pi^{(i)}(s)=(M_{b}+m_{q})^{2}M_{b}^{2}\sum_{j=-1}^{6}\sum_{k=0}^{3}%
A_{j,k}^{(i)}\left(  \ln\frac{-s}{M_{b}^{2}}\right)  ^{k}\left(  \frac
{M_{b}^{2}}{s}\right)  ^{j}\ \ \ \ \ (i=0,1,2).\label{PERT2}%
\end{equation}
and the coefficients $A_{j,k}^{(i)}$ of the QCD perturbative expansion are
given in \cite{CHETYRKINSTEINHAUSER} to order $(M_{b}^{2}/s)^{7}$. For
instance, the first term of the expansion is given by:
\begin{align*}
& \Pi^{(0)}(s)\,=\,\frac{3}{16\pi^{2}}(M_{b}+m_{q})^{2}s\left\{
3-2\ln\left(  \frac{-s}{M_{b}^{2}}\right)  +4\,\frac{M_{b}^{2}}{s}\ln\left(
\frac{-s}{M_{b}^{2}}\right)  +\right. \\
& \left[  -3-2\ln\left(  \frac{-s}{M_{b}^{2}}\right)  \right]  \left(
\frac{M_{b}^{2}}{s}\right)  ^{2}+\frac{2}{3}\left(  \frac{M_{b}^{2}}%
{s}\right)  ^{3}+\frac{1}{6}\left(  \frac{M_{b}^{2}}{s}\right)  ^{4}+\frac
{1}{15}\left(  \frac{M_{b}^{2}}{s}\right)  ^{5}+\\
& \left.  \frac{1}{30}\left(  \frac{M_{b}^{2}}{s}\right)  ^{6}+\frac{2}%
{105}\left(  \frac{M_{b}^{2}}{s}\right)  ^{7}+\ldots.\right\}
\end{align*}

In the asymptotic expansion (\ref{2QCD}) there are also non-perturbative terms
due to the quark and gluon condensates. In our calculations we will include
these terms up to dimension six \cite{REINDERS}:%
\begin{align}
\Pi^{\mathrm{nonpert.}}(s)  & =(M_{b}+m_{q})^{2}
\left\{
M_{b} \left\langle \bar {q}q \right\rangle 
\left[ 
\frac{1}{s-M_{b}^{2}}
\left( 1+ 2 \frac{\alpha_s}{\pi} \right) 
\right. 
\right.
\label{NONPERT}\\
&
\left.
 +  2 \frac{\alpha_s}{\pi} 
\ln \frac{M_{b}^{2}}{-s+ M_{b}^{2}} 
\right]  +O(m_{q}/M_{b})
\nonumber \\
& -\frac{1}{12}\frac{1}{s-M_{b}^{2}}
\left\langle \frac{\alpha_{s}} {\pi}G^{2}\right\rangle 
-\frac{1}{2}M_{b}
\left[  
\frac{1}{(s-M_{b}^{2})^{2}}+\frac{M_{b}^{2}}{(s-M_{b}^{2})^{3}}
\right]  
\left\langle \overline{q}\sigma Gq \right\rangle 
\nonumber\\
& 
\left.
-\frac{8}{27} \pi 
\left[  
\frac{2}{(s-M_{b}^{2})^{2}} + 
\frac{M_{b}^{2}}{(s-M_{b}^{2})^{3}}-\frac{M_{b}^{4}}{(s-M_{b}^{2})^{4}}
\right]
\alpha_{s} \left\langle \overline{q}q \right\rangle ^{2}
\right\}
\nonumber
\end{align}

The $\alpha_{s}$ correction to the quark condensate \cite{JAMINLANGE} turns
out to be small but non-negligible.

Before going on, a comment on the asymptotic expansion is in order. It is
known that the convergence of the perturbative expansion of the two-point
function, when written in terms of the pole mass, is rather poor. In fact, in
many calculations involving heavy quarks, the first and second order loop
contributions are typically of the same order of magnitude making difficult to
achieved convergence in the results. On the other hand, the expansion in terms
of the running mass converges much faster over a wide range of the
renormalization scale. Henceforward, for calculational purposes, we will
consider the relations among the pole and the running mass in the appropriate
order in the coupling constant \cite{KS,CHETYRKINSTEINHAUSER1,VERMASEREN} in
order to rewrite the perturbative piece of order $(\alpha_{s})^{i}$
(\ref{PERT2}) in the form
\begin{equation}
\Pi^{(i)}(s)=m_{b}^{2}(\mu)(m_{b}(\mu)+m_{q}(\mu))^{2}\sum_{j=-1}^{6}%
\sum_{k=0}^{3}\tilde{A}_{j,k}^{(i)}\left(  \ln\frac{-s}{\mu^{2}}\right)
^{k}\left(  \frac{m_{b}^{2}(\mu)}{s}\right)  ^{j}%
\ \ \ \ \ (i=0,1,2).\label{12}%
\end{equation}
and similarly for the non-perturbative piece. The coefficients $\tilde
{A}_{j,k}^{(i)}$ depend on the mass logarithms $\ln(m_{b}^{2}/\mu^{2})$
up to the third power. As $\Pi(s)$ is not known to all orders in $\alpha_{s}$,
the results of our analysis will depend to some extend on the choice of the
renormalization point $\mu$. In the sum rule considered here\ there are two
obvious choices, $\mu=m_{b}$ and $\mu=s_{0}$, the radius of the integration
contour. The former choice will sum up the mass logs of the form $\ln(m_{b}^{2}/\mu^2)$
and the latter choice the $\ln (-s/\mu^{2})$ terms. We find
that the for the choice $\mu=m_{b}$ the convergence of the perturbative terms
is significantly better, so we will adopt this value in the presentation here.

For the experimental side, we can split up the absorptive part into the
established lowest lying pseudoscalar $b\overline{q}$ resonance $B_{q}$ and an
unknown hadronic continuum :
\begin{equation}
\frac{1}{\pi}\,\,\mathrm{Im}\Pi(s)\,\,\,=\,\,\,M_{B_{q}}^{4}\,f_{B_{q}}%
^{2}\,\,\delta(s-M_{B_{q}}^{2})\,\,+\,\,\frac{1}{\pi}\,\,\mathrm{Im}%
\Pi^{\mathrm{cont.}}_{\mathrm{phys.}}\,\,\theta(s-s_{\mathrm{cont.}}^{\mathrm{phys.}})\label{IMAG1}%
\end{equation}
where $M_{\mathrm{B_{q}}}$ and $f_{\mathrm{B_{q}}}$ are, respectively, the
mass and the decay constants of the pseudoscalar meson $B_{q}$.

Looking back to equation (\ref{SR}) and taking all the theoretical parameters
including the mass of the $B_{q}$-meson as our inputs in the calculation, we
see that the decay constant could be computed if we would have accurate
information on the hadronic continuum contribution, which is, however, not the case.

To cope with this problem we make use of the freedom of choosing the
polynomial in equation (\ref{SR}). We take for $P(s)$ a polynomial of the
form:
\begin{equation}
P_{n}(s)=a_{0}+a_{1}s+a_{2}s^{2}+a_{3}s^{3}+\ldots+a_{n}s^{n},\label{POLY}%
\end{equation}
where the coefficients are fixed by imposing a normalization condition at
threshold 
\begin{equation}
P_{n} \left( s_{\mathrm{ph.}}=M_{B_{q}}^{2} \right) \, = \, 1,
\label{POLYNORM}
\end{equation}
and requiring that the
polynomial $P_{n}(s)$ should minimize the contribution of the continuum in the
range $\left[  s_{\mathrm{cont.}}^{\mathrm{phys.}},s_{0}\right]  $ in a least square sense,
i.e.,
\begin{equation}
\int_{s^{\mathrm{phys.}}_{\mathrm{cont.}}}^{s_{0}}s^{k}P_{n}(s)\,\,ds=0\,\,\mathrm{for}
\,\,k=0,\ldots n-1,
\label{POLY2}
\end{equation}
The polynomials obtained in this way are closely related to the Legendre polynomials.
In the appendix the explicit form of the set of polynomials use in this work is given.

This way of introducing the polynomial weight in the sum rule makes negligible 
the integration of 
$\frac{1}{\pi}\,\,\mathrm{Im}\,\Pi
^{\mathrm{cont.}}_{\mathrm{phys.}}\,\,\theta(s-s^{\mathrm{phys.}}_{\mathrm{cont.}})$,
so that in the phenomenological side of the sum rule only the contribution
of the $B_q$ resonance remains. On the other hand, as we will see in the results,
it increases the value of the duality parameter $s_0$ and, therefore, the asymptotic expansion of
QCD can be trusted safely in the integration contour.

To the extend that $\mathrm{Im}\,\,\Pi^{\mathrm{cont.}}_{\mathrm{phys.}}$ can be approximated by
an $n$ degree polynomial these conditions lead to an exact cancellation of the
continuum contribution to the left hand side of equation (\ref{SR}). As a
welcome side effect, this choice of polynomial will enhance the role of the
$B_{q}$ resonance. Notice however that increasing the degree of the polynomial
$P_{n}(s)$, will require the knowledge of further terms in the mass expansion
and in the non-perturbative series. Therefore the polynomial degree cannot be
chosen arbitrarily high.

To check the consistency of the method, we have considered a second to fifth
degree polynomials and the results are compatible within the range of the
errors introduced by the inputs of the calculation. We also have checked
explicitly that a smooth continuum contribution had no influence on the result.

The sum rule method above was previously used in the calculation of the charm
mass from the $c\overline{c}$ experimental data. The continuum data in this
case were known from the BES II collaboration \cite{bes} and were shown to
have no influence on the predicted mass \cite{PENARROCHASCHILCHER}. Employing
the same technique, a very accurate prediction of the bottom quark mass was
also obtained using the experimental information of the upsilon system
\cite{PENARROCHABORDESSCHILCHER}.

After these considerations we proceed with the analytical calculation. The
integrals that we have to evaluate on the right hand side of the sum rule,
equation (\ref{SR}), are
\begin{equation}
J(p,k)=\frac{1}{2\pi i}\oint_{\left\vert s\right\vert =s_{0}}s^{p}\left(
\ln\frac{-s}{\mu^{2}}\right)  ^{k}ds,\label{INTEG}%
\end{equation}
for $k=0,1,2,3$ and $p=-6,-5,..,n+1$. These integrals can be found e.g. in
reference \cite{PPSS}. After integration, equation (\ref{SR}) yields the sum
rule
\begin{align}
M_{B_{q}}^{4}\,\,f_{B_{q}}^{2}P(M_{B_{q}}^{2})  & = \, m_{b}^{2}(\mu)(m_{b}%
(\mu)+m_{q}(\mu))^{2}\label{SR2}\\
& \times\sum_{q=0}^{n}\sum_{i=0}^{2}\sum_{j=-1}^{6}\sum_{k=0}^{3}a_{q}\left(
\frac{\alpha_{s}(\mu)}{\pi}\right)  ^{i}\tilde{A}_{j,k}^{(i)}\,\,m_{b}%
^{2j}(\mu)\,\,J(q-j,k)\nonumber\\
& +\text{ non-perturbative terms}\nonumber
\end{align}
where, for brevity, we have not written down the non-perturbative terms
explicitly. The contribution of the continuum is neglected.

Plugging the theoretical and experimental inputs (physical threshold, quarks
and meson masses, condensates and strong coupling constant) into the sum rule,
we obtain the decay constant $f_{B_{q}}$ for various values of the degree $n$
of the polynomial and various values of $s_{0}$. Given the correct QCD asymptotic correlator and
the correct hadronic continuum, the calculation of the decay constant should,
of course, not depend either on $s_{0}$ or on the degree $n$ of the polynomial
in the sum rule (\ref{SR}). Accordingly, for a given $n$ we choose the flattest region
in the curve $f_B(s_0)$ to extract our prediction for the decay constant.
To be specific we choose the point of minimal slope. 
On the other hand, for different polynomials, the value of $f_B$, extracted in this way, could differ from
each other as the cancellation of the continuum may be incomplete or the QCD expansion
not accurate enough. We find, however, practically the same results for all our polynomials.
This additional stability is truly remarkable as the coefficients of the polynomials of
order $n=2,3,4$ and $5$ are completely different and the respective predictions are based on
completely different superpositions of finite energy moment sum rules. This extended
consistency leads us to attach great confidence in our numbers and associated errors.

\section{Results}

We calculate the decay constants for the $B$ and $B_{s}$ heavy mesons. In the
first case we take $m_{q}=0$ everywhere. In the second case we retain
$m_{q}=m_{s}\neq0$ in the factor $(M_{b}+m_{q})^{2}$ in front of the
correlation function only. Further terms in the power series in $m_{s}^{2}/s$
in (\ref{PERT1}) are completely negligible for the integration radii $s_{0}$
we use in the calculation.

The experimental and theoretical inputs are as follows. The physical threshold
$s_{\mathrm{ph.}}$ is the squared mass of the lowest lying resonance in the
$b\overline{q}$ channel. For $q$ being the light quark $u$, we have:%

\begin{equation}
s_{\mathrm{ph.}}=M_{B}^{2}=5.279^{2}\,\,\text{GeV}^{2}=27.87,\text{GeV}%
^{2}\label{BU}%
\end{equation}
whereas the continuum threshold $s^{\mathrm{phys.}}_{\mathrm{cont.}}$ is taken from the next
intermediate state $B\pi\pi$ in an s-wave $I=\frac{1}{2}$, i. e.
\[
s_{\mathrm{cont.}}^{\mathrm{phys.}}=\left(  M_{B}+2m_{\pi}\right)  ^{2}=30.90\,\,\mathrm{GeV}%
^{2}\,.
\]

For $q$ being the strange quark we take:
\begin{equation}
s_{\mathrm{ph.}}=M_{B_{s}}^{2}=5.369^{2}\,\,\text{GeV}^{2}=28.83,\text{GeV}%
^{2}\label{BS}%
\end{equation}
The continuum threshold starts in this case at the value:
\[
s_{\mathrm{cont.}}^{\mathrm{phys.}}=\left(  M_{B_{s}}+2m_{\pi}\right)  ^{2}%
=31.92\,\,\mathrm{GeV}^{2}.
\]

In the theoretical side of the sum rule we take the following inputs. The
strong coupling constant at the scale of the electroweak $Z$ boson mass
\cite{BETHKE}
\begin{equation}
\alpha_{s}(M_{Z})=0.118\pm0.003\label{ALPHA}%
\end{equation}
is run down to the computation scale using the four loop formulas of reference
\cite{SANTAMARIA}. For the quark and gluon condensates (see for example
\cite{JAMINLANGE}) and the mass of the strange quark \cite{PDG} we take:
\begin{align}
& <\overline{q}q>(2\,\mathrm{GeV})\,=\,(-267\,\pm\,17\,\,\mathrm{MeV})^{3},\nonumber\\
& <\frac{\alpha_{s}}{\pi}\,G\,G>\,=\,0.024\,\pm\,0.012\,\,\mathrm{GeV}^4,\nonumber\\
& <\overline{q}\sigma Gq>\,=\,m_{0}^{2}\,<\overline{q}q>,\,\,\,\mathrm{with}%
\,\,\,m_{0}^{2}\,=\,0.8\,\pm0.2\,\mathrm{GeV},\nonumber\\
& m_{s}(2 \,\mathrm{GeV})\,=\,120\,\pm50\,\mathrm{MeV},\nonumber\\
& <\overline{s}s>\,=\,(0.8\pm 0.3)<\overline{q}q>.\label{COND}%
\end{align}
As discussed, above we fix the renormalization scale $\mu$ in the theoretical
side of the sum to be $\mu=m_{b}(m_{b})$, as the perturbative series in
$\alpha_{s}$ is well under control, i.e. the first and second order terms are
only a few percent of the dominant zero order one and the second order one is
much smaller than the first order one. We use a reasonable variation of $\mu$
to estimate the corresponding error in our final result.

Finally, for the bottom quark, a value $m_{b}(m_{b})\approx4.20\,\mathrm{GeV}$
is generally accepted. For consistency we take the result $m_{b}%
(m_{b})\,=\,4.19\,\pm0.05\,\mathrm{GeV}$ of which has been obtained by us
\cite{PENARROCHABORDESSCHILCHER} with a similar sum rule method.

In order to calculate the decay constant for the pseudoscalar meson $B$, we
proceed in the way described above. We compute$\ f_{B}$ as a function of
$s_{0}$ with the four different sum rules (\ref{SR}) corresponding to
$n=2,3,4,5$. The results, plotted in Fig. 1 show remarkable stability
properties. We define the optimal value of $s_{0}$  as the center of the stability region 
(represented by a cross in Fig.1) where the first and/or second derivative of $f_B(s_0)$ vanishes.
At these values of $s_{0}$ we obtain the following consistent results:
\begin{align*}
f_{B}  & =175\text{ \ \ MeV  for \ \ }n=2\\
f_{B}  & =177\text{ \ \ MeV  for \ \ }n=3\\
f_{B}  & =178\text{ \ \ MeV  for \ \ }n=4\\
f_{B}  & =178\text{ \ \ MeV  for \ \ }n=5
\end{align*}

\begin{figure}[th]
\centerline{\psfig{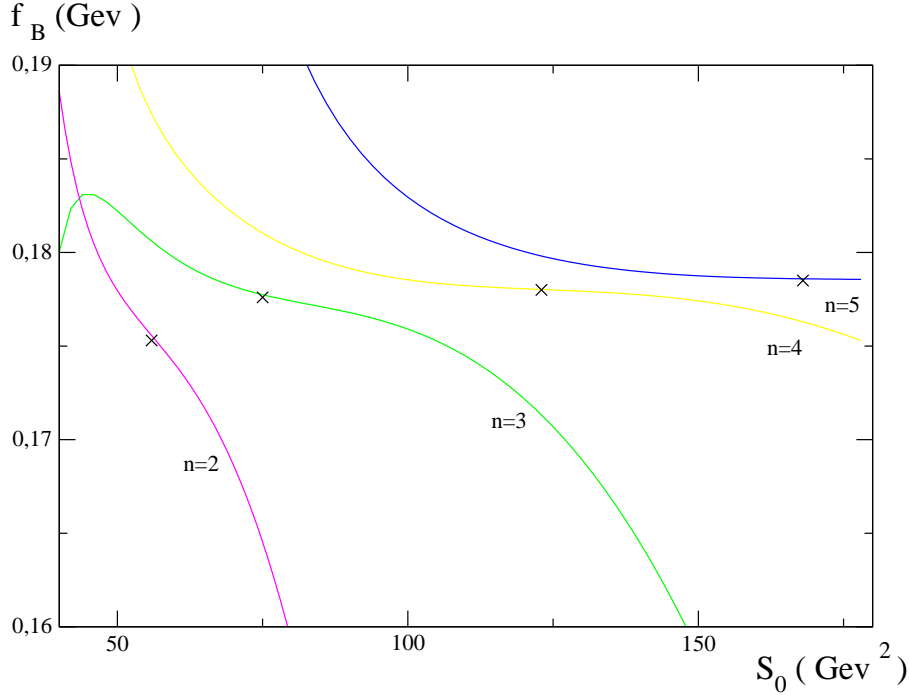}}
\caption{Decay constant $f_{B}$ as a function of the integration radius
$s_{0} $ for $m_{b}(m_{b})=4.19 \, \mathrm{GeV}$. The lines represent the
sum rule (\ref{SR}) for several choices of the polynomial $P_n(s)$.}%
\end{figure}

Notice from Fig. 1 that for the fourth degree polynomial (n=4) there is an stability region of about
$50 \, \, {\mathrm GeV}^2$ around $s_{0}$. In this region the decay constant changes by less than 
$1 \,$ percent. This stability region is pushed up to $100 \, \, {\mathrm GeV}^2$ in the case of $n=5$.
With these considerations we estimate a conservative
error inherent to the method of $\pm 3 \,  \mathrm{MeV}$. 

Other sources of errors arising in the calculation of $f_{B}$ are
the quark condensates which affect the result by $\pm3\,\mathrm{MeV,}$ and the
bottom mass which, in the range given above, produces a variation in the decay
constant of $-13,+10$ $\mathrm{MeV}$. This last one is the main source of
uncertainty in the final result. Finally we have changed the renormalization
scale in the range $\mu\,\in\,\left[  3,6\right]  \,\mathrm{GeV}$. We estimate
an error of $\pm6 \,\mathrm{MeV}$ associated to this change in $\mu$ which
is roghly related to the convergence of the asymptotic expansion 

Other errors due to the QCD side of
\ the sum rule (higher order terms in $m_{b}^{2}/s$ and the error on
$\alpha_{s}(m_{Z})$ are negligible in comparison).

Adding quadratically all the errors, we finally quote the following result for
the decay constant of the light meson $B$:
\begin{equation}
f_{B}\,=\,178\,\pm\,14\,(\mathrm{inp.})\,\pm\,3\,(\mathrm{meth.}%
)\,\,\,\mathrm{MeV}.\label{RESULTFB}%
\end{equation}

The first error comes from the inputs of the computation  and the truncated QCD theory whereas the second is
due to the method itself.

\begin{figure}[th]
\centerline{\psfig{figure=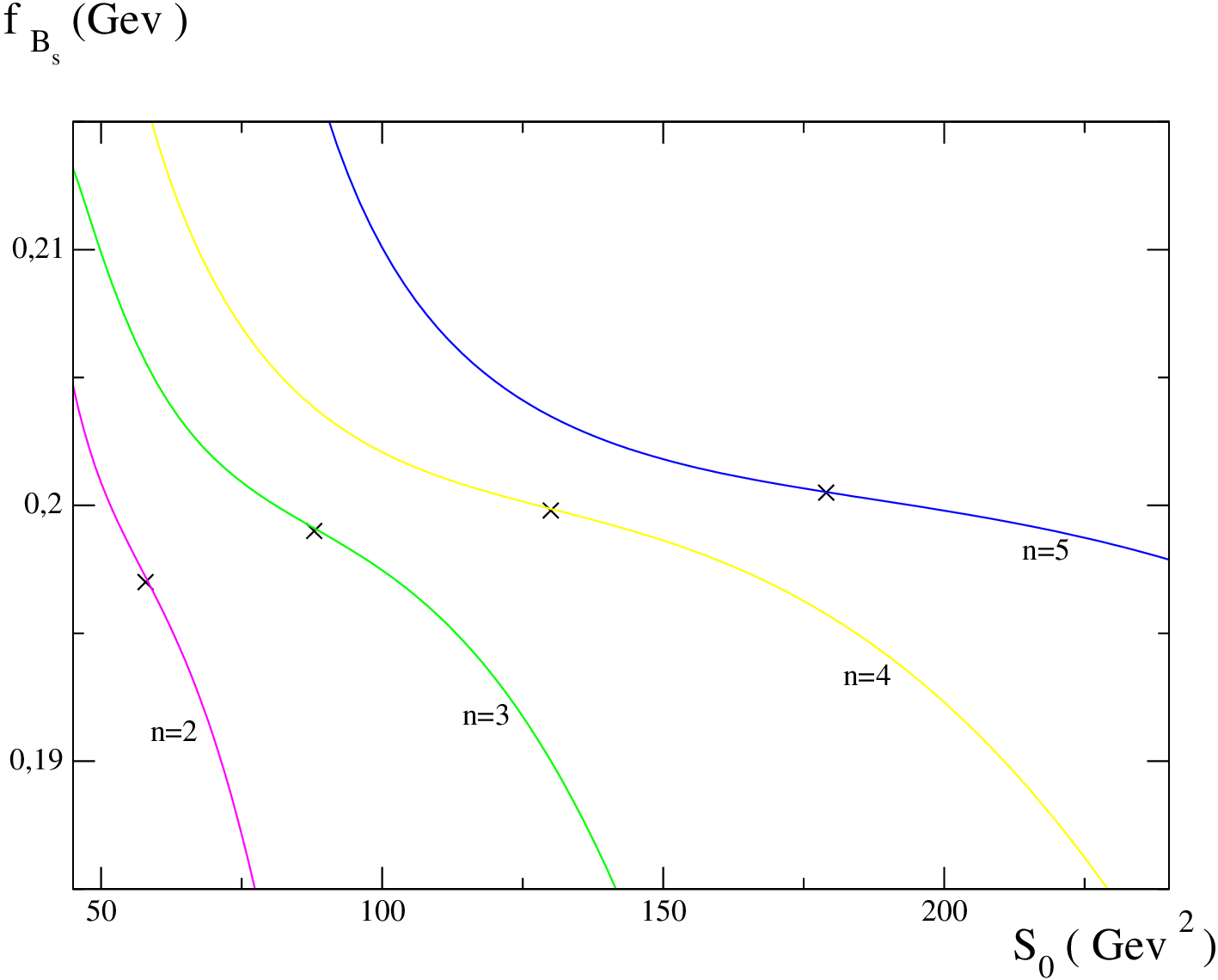,scale=0.5}}
\caption{Decay constant $f_{B_{s}}$ as a function of the integration radius
$s_{0}$ for $m_{b}(m_{b})=4.19\,\mathrm{GeV}$. The lines represent the
sum rule (\ref{SR}) for several choices of the polynomial $P_n(s)$.}%
\end{figure}
Proceeding in the same fashion, but keeping the mass of the strange quark in
the overall factor and the order $m_s/m_b$ in the one loop contribution, we find the decay constant for the $B_{s}$ meson
($f_{B_{s}}$),
\[
f_{B_{s}}\,=\,200\,\pm\,14\,(\mathrm{inp.})\,\pm\,3\,(\mathrm{meth.}%
)\,\,\,\mathrm{MeV}.
\]
(compare Fig. 2)

In the analysis of theoretical errors the only new ingredient is the
uncertainty coming from the strange quark mass which turns out to be negligible.

The ratio of the decay constants $f_{B_{s}}$ and $\ f_{B}$ (which would be $1
$ in the chiral limit) is of special interest. We find:%

\begin{equation}
\frac{f_{B_{s}}}{f_{B}}\,=\,1.12\,\pm\,0.01\,(\mathrm{inp.})\,\pm
\,0.03\,(\mathrm{meth.}).
\end{equation}

In the calculation, the uncertainties due to the method and to the theoretical
inputs (mainly to the bottom quark mass) are correlated, so that the final
error is very small.

Finally we should mention that in the past the stability region was often
determined in a rather ad hoc fashion by considering the intercept of sum rule
predictions for moments differing by one power. We find that, suitably
modified, this prescription also works in our sum rule method with similar
results but larger errors.

\section{Conclusions}

In this note we have computed the decay constant of $B_{q}$-mesons for $q$
either the strange or the $u$ or $d$ massless quarks. We have employed a
judicious combination of moments in QCD finite energy sum rules in order to
minimize the shortcomings of the available experimental data. On the
theoretical side of the pseudoscalar two-point function, we have used in the
perturbative piece an expansion up to three loops in the strong coupling
constant and up to order $\left(  m_{b}^{2}/s\right)  ^{7}$ in the mass
expansion and in the non-perturbative piece we considered condensates up to
dimension six including the $\alpha_{s}$ correction in the leading term.
Instead of the commonly adopted pole mass of the bottom quark, we use the
running mass to get good convergence of the perturbative series. It turns out
that for the renormalization point $\mu\,=\,m_{b}(m_{b})$ the first and second
order contribution in the strong coupling are term by term much smaller. This
good convergence is due to the summing up of the mass logarithms.

In the sum rule, the contour integration of the asymptotic part is performed
analytically. This particular fact differs from other computations based on
sum rules where the asymptotic QCD is integrated along a cut of the two-point
function starting at the pole mass squared. The latter way to proceed is
problematic when loop corrections are included and the complete analytical QCD
expression along the cut is not known. In this approach QCD has to be
extrapolated from low energy to high energy \cite{JAMINLANGE}. We also differ
from many other sum rule calculation of $f_{B}$ in that we do not require two
largely unrelated sum rules to determine a stability point via an intercept of
the curves $f_{B}(s_{0})$.

Our results are very sensitive to the value of the running mass, giving most
of the theoretical uncertainty. On the other hand they turn out to be very
stable against the variations of the other parameters, in particular the
renormalization scale and the integration radius $s_{0}$.

Comparing with other results in the literature, our results agree within the
error bars with the ones obtained using sum rule methods and with lattice
computations. When compared with the most recent numbers
\cite{JAMINLANGE,APE,UKQCD,PS}, however, our results are lower.

\ The interest in our method lies in the fact that it approaches the problem
from a different angle and, in our opinion, is less proned to systematic errors.

\bigskip

\section*{Appendix}

For convenience of the reader in this appendix we list the first few polynomials
emerging from relations (\ref{POLYNORM},\ref{POLY2}). From the second condition, namely 
(\ref{POLY2}), it is easy to realize that the set of polynomials $P_n(s)$ are n-degree 
orthogonal polynomials in the interval of the variable $s \, \in \, [s_{\rm cont.}^{\mathrm{phys.}},s_0]$. 
Then, leaving aside the normalization condition (\ref{POLYNORM}) that we take, for convenience,
in order to stress the contribution of the lowest lying resonance in the sum rule, they are 
related to the so-called Legendre polynomials (${\cal P}_n(x)$) in the interval of the
variable $x \, \in \, [-1,1]$. After including this normalization
condition, we can write more precisely:
\begin{equation}
P_n(s)  \, = \, \frac{{\cal P}_n \left( x(s) \right)}{{\cal P}_n\left( x(M^2_{B_q}) \right)}
\label{LEGENDRE}
\end{equation}
Where the variable $x(s)$ is:
$$
x(s) \, = \, \frac{2 s \, - \, (s_0 + s_{\rm cont.}^{\mathrm{phys.}})}{s_0 - s_{\rm cont.}^{\mathrm{phys.}}}
$$
which ranges in the required interval when $s \, \in \, [s_{\rm cont.}^{\mathrm{phys.}},s_0]$.

The explicit form of these polynomials is well known and can be found, for instance, in
\cite{SANSONE}. Nevertheless, for sake of completeness, we quote here the ones we have used
in the calculation.
\begin{align*}
& {\cal P}_2 \left( x(s) \right) \, = \, \frac{3}{2} (x^2-1),
\\
& {\cal P}_3 \left( x(s) \right) \, = \, \frac{1}{2} (5 x^3 - 3 x),
\\
& {\cal P}_4 \left( x(s) \right) \, = \, \frac{1}{8} (35 x^4 - 30 x^2 + 3),
\\
& {\cal P}_5 \left( x(s) \right) \, = \, \frac{1}{8} (63 x^5 - 70 x^3 + 15 x).
\end{align*}

Finally, in Fig. 3, and in order to appreciate how the dumping of the experimental physical continuum in the sum rule is expected,
we have plotted the form of the polynomials $P_n(s)$ for n=2,3,4,5 at the stability values of $s_0$ used
in the calculation of $f_{B}$.

\begin{figure}[bt]
\centerline{\psfig{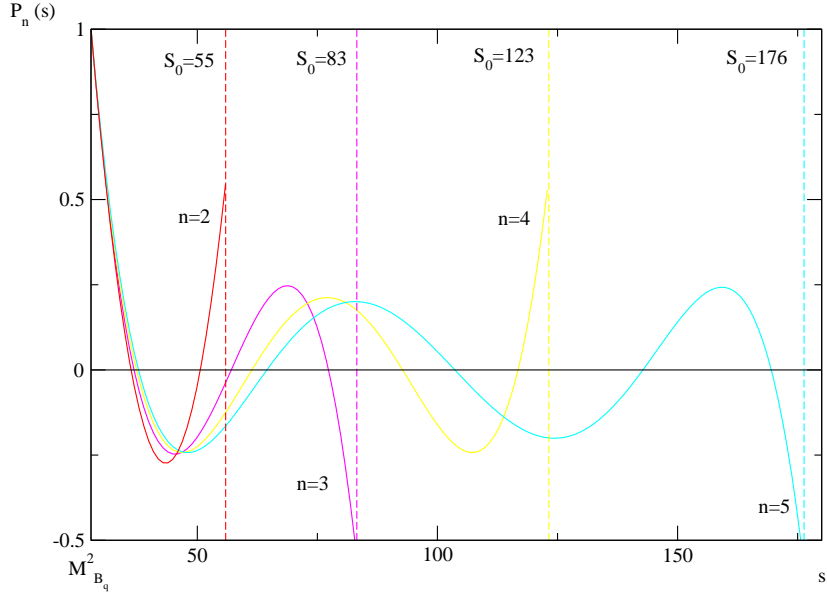}}
\caption{Polynomials $P_n(s)$ obtained from conditions (\ref{POLYNORM},\ref{POLY2})
taken at the stability values of $s_0$ in the calculation of $f_{B}$.}
\end{figure}

\end{document}